\documentclass[12pt,preprint]{aastex}

\shorttitle{The Araucaria Project: Distance to NGC~3109}
\shortauthors{Soszy{\'n}ski et al.}

\begin{document}

\title{The Araucaria Project: Distance to the Local Group Galaxy NGC~3109
from Near-Infrared Photometry of Cepheids\footnote{Based on observations
obtained with the ESO VLT for Large Programme 171.D-0004}}

\author{I. Soszy{\'n}ski}
\affil{Universidad de Concepci{\'o}n, Departamento de Fisica, Casilla 160--C,
  Concepci{\'o}n, Chile\\
  Warsaw University Observatory, Al.~Ujazdowskie~4, 00-478~Warszawa, Poland}
\email{soszynsk@astrouw.edu.pl}

\author{W. Gieren}
\affil{Universidad de Concepci{\'o}n, Departamento de Fisica, Casilla 160--C,
  Concepci{\'o}n, Chile}
\email{wgieren@astro-udec.cl}

\author{G. Pietrzy{\'n}ski}
\affil{Universidad de Concepci{\'o}n, Departamento de Fisica, Casilla 160--C,
  Concepci{\'o}n, Chile\\
  Warsaw University Observatory, Al.~Ujazdowskie~4, 00-478~Warszawa, Poland}
\email{pietrzyn@astro-udec.cl}

\author{F. Bresolin}
\affil{Institute for Astronomy, University of Hawaii at Manoa, 2680 Woodlawn
  Drive, Honolulu HI 96822, USA}
\email{bresolin@ifa.hawaii.edu}

\author{R.-P. Kudritzki}
\affil{Institute for Astronomy, University of Hawaii at Manoa, 2680 Woodlawn
  Drive, Honolulu HI 96822, USA}
\email{kud@ifa.hawaii.edu}

\and

\author{J. Storm}
\affil{Astrophysikalisches Institut Potsdam, An der Sternwarte 16, D-14482
  Potsdam, Germany}
\email{jstorm@aip.de}

\begin{abstract}
We present near-infrared $J$- and $K$-band photometry of 77 Cepheid
variables in the Local Group galaxy NGC~3109. Combining our data with the
previously published optical $V$- and $I$-band photometry of Cepheids in
this galaxy we derive an accurate distance and interstellar reddening to
NGC~3109. Adopting a distance modulus of 18.5 mag for the Large Magellanic
Cloud, we obtain a true distance modulus to NGC~3109 of
$(m-M)_0=25.571\pm0.024$~mag (random error), corresponding to a distance of
$1.30\pm0.02$ Mpc. The systematic uncertainty on this value (apart from the
adopted LMC distance) is of the order of $\pm$3\%, the main contributors to
this value being the uncertainty on the photometric zero points, and the
effect of blending with unresolved companion stars. The total reddening
determined from our multiwavelength solution is
$E(B-V)=0.087\pm0.012$~mag. About half of the reddening is produced
internal to NGC~3109. Our distance result is consistent with previous
determinations of the distance to NGC~3109, but has significantly reduced
error bars.
\end{abstract}

\section{Introduction}

The effectiveness of using multiwavelength optical and near-infrared (NIR)
observations of Cepheids for determining extragalactic distances has been
known for years \citep{mcg82,mf91}. However, only recently the technical
problems with obtaining reliable NIR photometry of faint objects in dense
regions have been solved. Using NIR photometry of the Cepheids provides a
number of advantages. First, the total and differential reddening is
significantly reduced in comparison with the optical bandpasses. Second,
the width of the Cepheid period--luminosity ({\it PL}) relation is
decreasing toward longer wavelengths. Third, NIR photometry of Cepheids is
thought to be less sensitive to eventual metallicity effects than optical
photometry. Fourth, the amplitudes of variability are significantly smaller
in the NIR than in the optical bands, so even single-epoch NIR observations
are sufficient to approximate the mean magnitudes. Moreover, \citet{sgp05}
showed that phase offsets and amplitude ratios between NIR and visual light
curves of the fundamental mode Cepheids are very stable. Thus, it is
possible to accurately transform random-phase single-epoch $JHK$
observations to mean magnitudes using the complete optical light curves.

Studying simultaneously the NIR and optical {\it PL} relations of Cepheids
provides one additional advantage. It allows to precisely determine the total
reddening, and measure the distance to nearby galaxies with an unprecedented
accuracy of better than 3\% \citep{gie05a,gie06,pie06a}.

NGC~3109 is a well-studied galaxy on the periphery of the Local Group. It
has been classified as an irregular galaxy \citep[Irr;][]{san61}, a dwarf
spiral \citep[Sm;][]{st81} or a dwarf barred spiral \citep[SBm;][]{dev91}.
The galaxy is seen almost edge-on at an inclination of $75^\circ\pm2$ and a
position angle of the disk of $93^\circ\pm2$ \citep{jc90}. The mean
metallicity of the old stellar population was estimated to be about $-1.7$
dex \citep{men02}. Little is known about the metallicity of the young
stellar population of NGC~3109. For one H II region in this galaxy
\citet{lgh03} have determined an oxygen abundance of about -1.0 dex, which
should be representative for the average metallicity of the Cepheids.

NGC~3109 is one of the largest and brightest Local Group galaxies visible
in the Southern hemisphere, and as such it was added to the list of targets
observed in the course of the Araucaria Project \citep{gie05b}. The project
aims at determining accurate distances to a number of nearby galaxies and
investigating the environmental dependences of a number of stellar
indicators: Cepheids, RR Lyrae stars, blue supergiants, the tip of the red
giant branch and red clump stars.

Historically, the first attempt of a Cepheid distance determination to
NGC~3109 was made by \citet{dik85}, who used photographic observations to
discover 5 Cepheids in this galaxy. They obtained a true distance modulus
of $(m-M)_0=25.98\pm0.15$~mag. This value, and all other distance
estimations presented in this paper, is tied to an assumed Large Magellanic
Cloud (LMC) distance modulus of 18.50~mag. \citet{sc88} increased the
number of known Cepheids in NGC~3109 to 29, and suggested the same distance
modulus to NGC~3109. Subsequent CCD photometry \citep{cpb92} revealed
systematic errors in the previous photographic data, and adjusted the
distance modulus estimation to $(m-M)_0=25.5\pm0.2$~mag. This revision
corresponded to a distance about 25\% shorter than the one obtained by
\citet{dik85} and \citet{sc88}. \citet{mpc97} increased the number of known
Cepheids in NGC~3109 by another 16 objects and secured multiwavelength
({\it BVRI}) photometry for some of them. They obtained a true distance
modulus of $(m-M)_0=25.67\pm0.16$~mag, but they adopted a total
interstellar reddening $E(B-V)=0$, because their forced solution yielded a
negative reddening value.

Recently, \citet[hereafter Paper I]{pie06b} presented a catalog of 113
Cepheids in NGC~3109 discovered from $V$ and $I$ observations collected with
the 1.3-m Warsaw Telescope at Las Campanas Observatory. 76 of these objects
were not known before, and for the remaining 37 Cepheids improved periods
were measured. Adopting a total reddening $E(B-V)=0.1$~mag a true distance
modulus $(m-M)_0=25.54\pm0.05$~mag (statistical error) was obtained.

In this work, we extend the light curve coverage for 77 of the Cepheids
presented in Paper I to the NIR $J$ and $K$ bands. We utilize multiband
NIR/optical photometry for an accurate determination of the distance, and
the total (average) interstellar extinction to the Cepheids in NGC~3109.

The paper is composed as follows. In Section 2, we describe the $J$ and $K$
band observations, data reductions and calibration of the photometry. In
Sections 3 and 4, we derive the Cepheid {\it PL} relations from our data and
determine the true distance modulus to NGC~3109 with the mean color
excess. Our results are discussed and summarized in Sections 5 and 6.

\section{Observations and Data Reduction}

We used deep $J$- and $K$-band images recorded with the 8.2-m ESO Very Large
Telescope equipped with the Infrared Spectrometer And Array Camera
(ISAAC). Fig.~1 shows the location of the three $2\makebox[0pt][l]{.}'5\times
2\makebox[0pt][l]{.}'5$ fields observed in service mode on 6 nights between 31
Jan and 19 Feb 2004. Each field was observed in $J$ and $K$ bands two times
on two different nights, with exception of the field F-I, for which data in the
$J$ band was only obtained once. Observations were carried out using a
jitter imaging technique, with a dithering of the frames following a random
pattern characterized by typical offsets about $10'$. The final frames in the
$J$ and $K$ bands were obtained as a co-addition of 32 and 88 single exposures
obtained with integration times 30~s and 15~s, respectively. Thus, the total
exposure time for a given observation was 16 minutes in $J$ and 22 minutes in
$K$. The observations were obtained under very good seeing conditions. During
four of six observing nights photometric standard stars on the UKIRT system
\citep{haw01} were observed along with the science fields.

The images were reduced using the program {\sc Jitter} from the {\sc Eclipse}
package developed by ESO to reduce the NIR data. The PSF photometry was
obtained with the programs {\sc Daophot} and {\sc Allstar}. The PSF model was
derived iteratively from 20-30 isolated bright stars following the procedure
described by \citet{pgu02}. In order to convert our profile photometry to the
aperture system, aperture corrections were computed using the same stars as
for the calculation of the PSF model. The median of the aperture corrections
obtained for all the stars was finally adopted as the aperture correction for
a given frame. The aperture photometry for our standard stars was performed
with {\sc Daophot} using the same aperture as for the calculation of the
aperture corrections. The photometry obtained on the two nights with no
standard observations was tied to the standard system using several hundred
comparison stars.

The astrometric solution for the observed fields was performed by
cross-identification of the brightest stars in each field with the Infrared
Digitized Sky Survey 2 (DSS2-infrared) images. We used programs developed by
\citet{uda98} to calculate the transformations between pixel grid of our
images and equatorial coordinates of the DSS astrometric system. The internal
error of the transformation is less than 0.3 arcsec, but systematic errors of
the DSS coordinates can be up to about 0.7 arcsec.

We performed an external check of our photometry by comparing the magnitudes
of the brightest stars ($K<16$ mag, $J<17$ mag) with the 2MASS Point Source
Catalog \citep{cut03}. Unfortunately, even the most luminous stars in our
dataset are close to the limiting magnitudes of the 2MASS catalog. Only about
one dozen 2MASS stars identified with our objects have photometric errors
smaller than 0.1 mag. However, for these stars we have not noticed any evident
zero point offsets between both datasets. We estimate that the agreement
between the zero points of both photometries is better than 0.03 mag.

Our three fields in the central regions of NGC~3109 contain 77 of the 113
Cepheids listed in Paper I. All the individual observations in $K$ and $J$ are
given in Table~1, which lists the stars' IDs, Heliocentric Julian Day of the
observations, and measurements in $K$ and $J$ with the standard
deviations. For most of the stars we collected two observations per given
filter. The exception are $J$-band observations of stars in the field F-I for
which only one point was secured. There are some other objects for which we
obtained only one observation in the $J$ or $K$ band. For most of these cases
the cause was the location of the variable close to the edge of the field and
imperfections of the telescope pointing. On the other hand, there are three
variables for which we collected more than two data points because they were
located in the overlapping parts of adjacent fields.

\section{The Cepheid Period--Luminosity Relations in $J$ and $K$}

All the individual $J$ and $K$ measurements reported in Table~1 were
transformed to the mean magnitudes of the Cepheids using the recipe given by
\citet{sgp05}. The corrections were derived using the complete $V$-band light
curves from Paper~I. For the vast majority of our Cepheids the mean magnitudes
obtained from the independent measurements at different phases agreed very
well, especially for objects discovered during the earlier surveys
\citep{sc88,mpc97} for which improved periods were available. The precisely
determined $V$-band phases are of crucial importance for the accuracy of the
estimated mean $K$ and $J$ magnitudes. For most of our objects, the difference
between the two estimates of the mean magnitude were comparable with the
measurement errors of the original points.

Table~2 gives the intensity mean $J$ and $K$ magnitudes of our sample of
Cepheids. Each value was derived as an average from the individual
determinations of the mean luminosities. In Table~2, we also provide the
periods (from Paper~I), uncertainties on the mean magnitudes (which contain
the intrinsic error 0.03 mag of the mean magnitude estimation technique) and
remarks on some of the variables.

In Fig.~2, we display the $J$- and $K$-band {\it PL} diagrams for the Cepheids
in NGC~3109. Following Paper I, the Cepheids with $\log{P}<0.75$ were excluded
from our distance determination. In this range of periods the scatter of
points around the average {\it PL} relation substantially grows, which is an
effect of the larger photometric errors and the contamination with probable
first overtone pulsators. It is also clearly appreciated from Fig.~2 that
below our adopted cutoff period a Malmquist bias begins to appear, due to the
depth limit of our photometry. Especially in the K band, for smaller periods
we see only the Cepheids lying towards the bright end of the instability strip
at these periods, whereas the fainter Cepheids are mostly below the detection
threshold.

Ideally, we should retain the same value of the cutoff period for all our
target galaxies. However, in practice this is not the best solution because
our photometry extends to different levels of faintness in the different
galaxies our program, corresponding to different periods at which an
incompleteness bias begins to be a serious problem. Also, there is quite a
variety in the relative numbers of long- and short-period Cepheids in the
different galaxies, making the choice of a uniform value of the cutoff period
difficult. We can, however, check on the effect of changing the cutoff period
to larger values. Our present solution for the distance modulus of NGC~3109
(see below) changes by less than 0.01 mag if the cutoff period of
$\log{P}=0.75$ is changed to values of 0.80, 0.90, and 1.0. In the $J$ band,
the maximum change is 0.04~mag, or 2\%, which is 1~$\sigma$, and therefore not
significant. We therefore conclude that our distance result for NGC~3109 does
not depend in any significant way on the adopted cutoff period (as long as its
value is large enough to avoid the problem of Malmquist bias and contamination
with overtone Cepheids).

We also omitted 5 variables with periods longer than our adopted cutoff
period, but significantly brighter than the ridge line luminosity at the
respective periods. All these outliers from the NIR {\it PL} relation are also
overluminous in the $V$ and $I$ bands (Paper I), though, in some cases, the
NIR deviation is larger than in the visual bands. A possible explanation of
this behavior is the presence of a bright red companion star (but see our
discussion). All the excluded objects are marked with empty circles in Fig.~2.

The straight lines superimposed in Fig.~2 are the best-fitting linear
functions with slopes adopted from the LMC Cepheids as given by
\citet{per04}: $-3.153\pm0.051$ and $-3.261\pm0.042$ for $J$ and $K$ bands,
respectively. Free least-square fits to the {\it PL} relations yield
somewhat shallower slopes: $-2.85\pm0.12$ for $J$ and $-3.11\pm0.12$ for
the $K$-band, but still statistically consistent with the Persson et
al. slopes. The weighted least-squares fits with forced slopes yield the
following relations:

$$J=-3.153\log{P}+23.452(\pm 0.028),~~~\sigma=0.178$$
$$K=-3.261\log{P}+23.123(\pm 0.027),~~~\sigma=0.181$$

To determine the relative distance moduli between NGC~3109 and LMC we need
to convert the NICMOS (LCO) photometric system used by \citet{per04} to the
UKIRT system utilized in this paper. According to Hawarden et al. (2001),
the magnitudes in both systems differ by constant, color-independent
values: $0.034\pm0.004$ and $0.015\pm0.007$ for the $J$ and $K$ wavebands,
respectively. Applying these offsets, we obtained the following relative
apparent distance moduli with respect to the LMC:
$\delta(m-M)_J=7.150\pm0.028$~mag and $\delta(m-M)_K=7.102\pm0.027$~mag,
or, assuming the LMC distance modulus of 18.5~mag,
$(m-M)_J=25.650\pm0.028$~mag and $(m-M)_K=25.602\pm0.027$~mag.

Taking into consideration the apparent distance moduli measured in the $V$
and $I$ bands in Paper I, $(m-M)_V=25.854\pm0.027$~mag and
$(m-M)_I=25.739\pm0.026$~mag, we can determine the true distance modulus
and total interstellar reddening to NGC~3109. In Fig.~3 we present the
apparent distance moduli for the $V$, $I$, $J$ and $K$ wavebands plotted
against the total-to-selective absorption provided by the \citet{sfd98}
reddening law. One can notice the extraordinary agreement between the
measured distance moduli and the linear function fitted to the points. The
slope and the intersection of this relation give $E(B-V)$ and the true
distance modulus to the galaxy, respectively. The best-fitting relation
yields:

$$E(B-V)=0.087\pm0.012\ \mathrm{mag}$$
$$(m-M)_0=25.571\pm0.024\ \mathrm{mag}$$

\noindent corresponding to a distance 1.30 Mpc $\pm$ 0.02 Mpc.

\section{Period -- NIR Wesenheit Index Relation}

The Wesenheit index \citep{mad82} is a reddening-free quantity defined as a
linear combination of the selected magnitude and color of the star. For
example, for $J$ and $K$ magnitudes it is defined as:

$$W_{JK}=K-\frac{A_K}{E(J-K)}(J-K)$$

Using Schlegel's et al. (1998) ratios of total-to-selective absorption
measured for the UKIRT system ($R_J=0.902$, $R_K=0.367$), we obtain
$A_K/E(J-K)=0.686$. The advantage of using the Wesenheit index is the
canceling of the interstellar extinction effect star by star without
explicitly determining this effect.

Fig.~4 shows the period--$W_{JK}$ diagram for our sample of Cepheids in
NGC~3109. The straight line shows the linear least-squares fit to the selected
Cepheids indicated with filled circles. Similarly to the procedure adopted for
the $J$ and $K$ bands, we force the slope of the $\log{P}$--$W_{JK}$ relation
to the relation defined by the LMC Cepheids \citep{per04}. After converting
Persson's et al. photometry onto the UKIRT system, we obtained the following
period--$W_{JK}$ relation for the LMC Cepheids:

$$W_{JK}=-3.374\log{P}+15.865$$

Solving for the best coefficients of the period--$W_{JK}$ relation in NGC~3109
leads to the following relation:

$$W_{JK}=-3.374\log{P}+22.944 (\pm0.030)$$

\noindent which corresponds to a NGC~3109 distance modulus of 25.579~mag.
This value is fully consistent with the true distance modulus obtained from
the multiwavelength analysis.

\section{Discussion}

An exhaustive discussion about possible systematic errors that can affect our
distance determination was presented by \citet{gie05a,gie06} and
\citet{pie06a}. Here we describe only the most important issues concerning
this subject.

The source of largest systematic error on our distance determination is
probably the unsolved problem of the LMC distance modulus. The possible
uncertainty on the distance to the LMC may exceed 10\%. For consistency with
our previous work, as well as with many other extragalactic distance
determinations, we assumed that the true distance modulus to the LMC is equal
to 18.50~mag. If future studies change this determination our distance
moduli can be easily transformed to the proper value.

Another potential source of systematic uncertainty of our results is the
unknown effect of metal abundances on the slopes and zero points of Cepheid
{\it PL} relations. To date very few empirical studies deal with this
problem, especially in the NIR domain. Theoretical considerations are also
inconclusive. Linear pulsation models \citep[e.g.][]{sg98,sbt99,ali99,ba01}
suggest that the dependence of the Cepheid {\it PL} relation on chemical
composition is very weak. On the other hand, the nonlinear models
\citep[e.g.][]{bon99,cmm00} predict that metal-rich Cepheids are
significantly fainter than metal-poor ones, but the effect decreases with
increasing wavelength. In our previous studies \citep{gie05a,gie06,pie06a}
we have not noticed any statistically significant relationships between
slopes of the NIR {\it PL} relations and mean metallicities. In the case of
NGC~3109 the {\it PL} relation seem somewhat shallower than in the LMC, but
taking into account the relatively smaller range of periods, the
disagreement is statistically insignificant. Our knowledge about the
influence of metallicity on the zero points of the NIR {\it PL} relations
is even more incomplete. While preliminary results from our project seem to
indicate that the effect of metallicity on the {\it PL} relation zero point
is very modest \citep{pg02}, we will have a much better database to
investigate the effect once we have measured the distances to all our
target galaxies with a variety of methods, including red clump stars
\citep{pgu03}, and the blue supergiant Flux-Weighted Gravity-Luminosity
Relation \citep{kbp03}. We will therefore leave an exhaustive discussion of
this point to a later stage of the Araucaria Project, and for the time
being {\it assume} that the Cepheid {\it PL} relation is universal.

Selection effects, such as an inhomogeneous distribution of the Cepheids
in the instability strip, do not influence significantly the total
error because our Cepheid sample is large enough to minimize such random
effects. Our variables seem to be randomly distributed across the strip, with
no tendency for grouping near the red or blue edges.

Similar conclusions can be drawn for possible crowding effects. Recently,
\citet{bre05} demonstrated that blending in the Sculptor galaxy NGC~300
affects the distance determination from ground-based Cepheid photometry by
less than 2\%. In the case of NGC~3109, the effect should be even smaller
because the galaxy is much closer, and the average density of stars is smaller.

The group of five overluminous stars in the period--luminosity diagrams which
were excluded in the distance determination are potentially interesting
objects. While one possibility for their excessive brightness is that they
are normal Cepheids which are blended with very bright unresolved stars, it is
intriguing that {\it all} these objects are about 1 mag brighter than the PL
relation at the corresponding period. If blending was the cause for the
over-brightness of these stars, one might expect a continuum of luminosity
offsets, and not the same value for each of them. This might suggest that an
{\it intrinsic} cause is responsible for these objects to be so luminous. If
these objects are binary Cepheids, they would certainly harbour some very
interesting new information on binary star formation for massive stars. Such
an investigation is beyond the scope of this work, however. We just mention
that very similar, overluminous Cepheids are also seen in IC 1613
\citep{pie06a}, and NGC 6822 \citep{gie06}, making it worthwhile to look more
closely into the nature of these stars.

From this discussion, and from the conclusions presented in the previous
papers of this series we conclude that the total systematic error on our
distance determination does not exceed $\sim$3\%. However, one should remember
that this estimate does not contain the currently largest uncertainty -- the
still debated distance modulus to the LMC.

Our determination of the total color excess ($E(B-V)=0.087$~mag), based on the
multiwavelength approach, can be compared with the Galactic reddening maps of
\citet{sfd98}. According to these maps, the foreground $E(B-V)$ changes across
NGC 3109 from 0.03 to 0.07 mag. Our estimation then indicates that the {\it
intrinsic} average reddening in NGC 3109 appropriate to the Cepheids is about
0.04 mag.

\section{Summary and Conclusions}

All previous CCD surveys for Cepheids in NGC~3109 were carried out at blue
wavelengths, thus the effect of obscuration by interstellar dust was a
significant factor affecting the final results. In the present study, we
present for the first time NIR CCD photometry of the Cepheids in this galaxy,
and we have determined an accurate distance and reddening. Although NGC~3109
does not contain a population of very long-period Cepheids, which carry the
strongest weight in the distance determinations, we were able to measure the
distance to this galaxy relative to the LMC with an accuracy of about 3\%.

In Table 3, we list the most important previous distance determinations to
NGC~3109. Fig.~5 shows the same results in a different way, highlighting the
trend of increasing accuracy in the results over the past two decades. It can
be appreciated that our result derived in this paper is in excellent agreement
with the most recent estimations of the distance to NGC~3109, but it is
clearly more accurate and will therefore be very useful for an improved
determination of environmental effects on stellar distance indicators. Such
studies will be the subject of forthcoming papers which will take advantage of
the series of accurate Cepheid distances to nearby galaxies from NIR
photometry which we are providing in our project.

\acknowledgments
{WG and GP gratefully acknowledge financial support for this work from the
Chilean Center for Astrophysics FONDAP 15010001. Support from the Polish KBN
grant No 2P03D02123 and BST grant to Warsaw University Observatory is also
acknowledged. We thank the ESO OPC for the generous amounts of observing time
at Paranal and La Silla telescopes allocated to our Large Programme. Special
thanks goes to the Paranal staff astronomers for their expert help in
obtaining the high-quality data which form the basis for the work reported in
this paper.}

\clearpage

\begin{deluxetable}{lcccccc}
\tablecolumns{7}
\tablewidth{0pc}
\tablecaption{Journal of the individual $J$ and $K$ observations of the NGC~3109 Cepheids.}
\tablehead{
\multicolumn{1}{c}{ID} & $J$ HJD & $J$ & $\sigma_J$ & $K$ HJD & $K$ & $\sigma_K$}
\startdata
cep001 &       --      &   --   &  --   & 2453036.80998 & 18.392 & 0.020 \\
cep001 & 2453047.68199 & 18.730 & 0.014 & 2453047.76282 & 18.265 & 0.018 \\
cep002 & 2453037.69282 & 18.966 & 0.021 & 2453037.76573 & 18.374 & 0.025 \\
cep002 & 2453049.67836 & 18.730 & 0.041 & 2453049.74253 & 18.165 & 0.016 \\
cep003 &       --      &   --   &  --   & 2453036.80998 & 18.632 & 0.026 \\
cep003 & 2453047.68199 & 19.017 & 0.024 & 2453047.76282 & 18.495 & 0.026 \\
cep004 &       --      &   --   &  --   & 2453036.80998 & 18.643 & 0.023 \\
cep004 & 2453047.68199 & 18.882 & 0.014 & 2453047.76282 & 18.394 & 0.019 \\
cep005 & 2453037.69282 & 19.338 & 0.030 & 2453037.76573 & 18.831 & 0.039 \\
cep005 & 2453049.67836 & 18.899 & 0.030 & 2453049.74253 & 18.352 & 0.027 \\
cep007 & 2453045.77048 & 19.873 & 0.023 & 2453045.83694 & 19.255 & 0.035 \\
cep007 & 2453055.62418 & 19.715 & 0.031 & 2453055.70117 & 19.237 & 0.029 \\
cep009 & 2453045.77048 & 19.408 & 0.022 & 2453045.83694 & 18.829 & 0.024 \\
cep009 & 2453055.62418 & 19.231 & 0.032 & 2453055.70117 & 18.865 & 0.021 \\
cep011 & 2453045.77048 & 19.874 & 0.026 & 2453045.83694 & 19.342 & 0.037 \\
cep011 & 2453055.62418 & 19.649 & 0.032 & 2453055.70117 & 19.121 & 0.027 \\
cep012 & 2453045.77048 & 19.739 & 0.029 & 2453045.83694 & 19.286 & 0.048 \\
cep012 & 2453055.62418 & 19.938 & 0.048 & 2453055.70117 & 19.338 & 0.039 \\
cep014 & 2453037.69282 & 19.616 & 0.041 & 2453037.76573 & 19.236 & 0.042 \\
cep014 & 2453045.77048 & 19.486 & 0.026 & 2453045.83694 & 19.202 & 0.036 \\
cep014 & 2453049.67836 & 19.777 & 0.057 & 2453049.74253 & 19.221 & 0.035 \\
cep014 & 2453055.62418 & 19.764 & 0.051 & 2453055.70117 & 19.448 & 0.032 \\
cep015 &       --      &   --   &  --   & 2453036.80998 & 19.250 & 0.036 \\
cep015 & 2453047.68199 & 19.847 & 0.027 & 2453047.76282 & 19.334 & 0.032 \\
cep016 & 2453037.69282 & 19.540 & 0.025 & 2453037.76573 & 19.315 & 0.038 \\
cep016 & 2453049.67836 & 19.800 & 0.039 & 2453049.74253 & 19.291 & 0.024 \\
cep017 & 2453045.77048 & 19.789 & 0.024 & 2453045.83694 & 19.353 & 0.035 \\
cep017 & 2453055.62418 & 20.061 & 0.039 & 2453055.70117 & 19.590 & 0.032 \\
cep018 & 2453037.69282 & 20.078 & 0.043 & 2453037.76573 & 19.597 & 0.041 \\
cep018 & 2453049.67836 & 20.076 & 0.072 &       --      &   --   &  --   \\
cep020 & 2453045.77048 & 20.250 & 0.029 & 2453045.83694 & 19.852 & 0.048 \\
cep020 & 2453055.62418 & 20.250 & 0.036 & 2453055.70117 & 19.639 & 0.033 \\
cep022 & 2453045.77048 & 19.907 & 0.038 & 2453045.83694 & 19.501 & 0.052 \\
cep022 & 2453055.62418 & 19.775 & 0.057 & 2453055.70117 & 19.479 & 0.037 \\
cep023 & 2453045.77048 & 19.625 & 0.021 & 2453045.83694 & 19.032 & 0.029 \\
cep023 & 2453055.62418 & 19.553 & 0.044 & 2453055.70117 & 19.142 & 0.025 \\
cep025 & 2453045.77048 & 18.638 & 0.018 & 2453045.83694 & 17.826 & 0.016 \\
cep025 & 2453055.62418 & 18.695 & 0.027 & 2453055.70117 & 17.803 & 0.019 \\
cep026 & 2453037.69282 & 20.298 & 0.045 & 2453037.76573 & 19.874 & 0.053 \\
cep027 & 2453045.77048 & 20.125 & 0.029 & 2453045.83694 & 19.634 & 0.058 \\
cep027 & 2453055.62418 & 20.188 & 0.052 & 2453055.70117 & 19.776 & 0.035 \\
cep028 &       --      &   --   &  --   & 2453036.80998 & 19.561 & 0.036 \\
cep028 & 2453047.68199 & 20.206 & 0.038 & 2453047.76282 & 19.688 & 0.051 \\
cep029 & 2453037.69282 & 19.880 & 0.035 & 2453037.76573 & 19.388 & 0.042 \\
cep029 & 2453049.67836 & 19.818 & 0.042 & 2453049.74253 & 19.276 & 0.024 \\
cep030 & 2453037.69282 & 20.397 & 0.048 & 2453037.76573 & 20.043 & 0.044 \\
cep030 & 2453049.67836 & 20.321 & 0.081 & 2453049.74253 & 19.789 & 0.037 \\
cep031 & 2453045.77048 & 19.934 & 0.028 & 2453045.83694 & 19.452 & 0.044 \\
cep031 & 2453055.62418 & 20.020 & 0.035 & 2453055.70117 & 19.463 & 0.030 \\
cep032 & 2453037.69282 & 20.479 & 0.043 & 2453037.76573 & 19.666 & 0.062 \\
cep032 & 2453049.67836 & 20.351 & 0.055 & 2453049.74253 & 19.529 & 0.059 \\
cep035 & 2453047.68199 & 20.043 & 0.050 & 2453047.76282 & 19.720 & 0.059 \\
cep036 & 2453037.69282 & 20.182 & 0.053 & 2453037.76573 & 19.811 & 0.049 \\
cep036 & 2453049.67836 & 20.266 & 0.076 & 2453049.74253 & 19.997 & 0.043 \\
cep038 &       --      &   --   &  --   & 2453036.80998 & 19.990 & 0.043 \\
cep038 & 2453047.68199 & 20.392 & 0.044 & 2453047.76282 & 19.926 & 0.054 \\
cep039 & 2453045.77048 & 20.609 & 0.047 & 2453045.83694 & 20.229 & 0.067 \\
cep039 & 2453055.62418 & 20.515 & 0.068 & 2453055.70117 & 20.201 & 0.043 \\
cep043 &       --      &   --   &  --   & 2453036.80998 & 20.105 & 0.058 \\
cep043 & 2453047.68199 & 20.409 & 0.042 & 2453047.76282 & 20.392 & 0.084 \\
cep047 &       --      &   --   &  --   & 2453036.80998 & 19.737 & 0.043 \\
cep047 & 2453047.68199 & 20.401 & 0.048 & 2453047.76282 & 19.948 & 0.066 \\
cep048 & 2453037.69282 & 20.586 & 0.038 & 2453037.76573 & 20.256 & 0.059 \\
cep048 & 2453049.67836 & 20.468 & 0.073 & 2453049.74253 & 19.722 & 0.033 \\
cep050 &       --      &   --   &  --   & 2453036.80998 & 19.955 & 0.042 \\
cep050 & 2453047.68199 & 20.707 & 0.064 & 2453047.76282 & 20.305 & 0.068 \\
cep051 & 2453045.77048 & 20.549 & 0.038 & 2453045.83694 & 20.518 & 0.076 \\
cep051 & 2453055.62418 & 20.462 & 0.055 & 2453055.70117 & 20.054 & 0.042 \\
cep052 &       --      &   --   &  --   & 2453036.80998 & 20.153 & 0.055 \\
cep052 & 2453047.68199 & 20.530 & 0.049 & 2453047.76282 & 20.215 & 0.058 \\
cep053 &       --      &   --   &  --   & 2453036.80998 & 20.020 & 0.048 \\
cep053 & 2453047.68199 & 20.816 & 0.054 & 2453047.76282 & 20.532 & 0.086 \\
cep054 &       --      &   --   &  --   & 2453036.80998 & 20.002 & 0.047 \\
cep054 & 2453047.68199 & 20.533 & 0.044 & 2453047.76282 & 20.340 & 0.072 \\
cep055 &       --      &   --   &  --   & 2453036.80998 & 19.929 & 0.061 \\
cep055 & 2453047.68199 & 20.695 & 0.059 & 2453047.76282 & 20.199 & 0.064 \\
cep056 & 2453045.77048 & 19.782 & 0.037 & 2453045.83694 & 18.693 & 0.027 \\
cep056 & 2453055.62418 & 20.012 & 0.039 & 2453055.70117 & 18.735 & 0.048 \\
cep057 & 2453037.69282 & 19.957 & 0.038 & 2453037.76573 & 19.021 & 0.067 \\
cep057 & 2453049.67836 & 19.571 & 0.051 & 2453049.74253 & 18.810 & 0.069 \\
cep058 & 2453045.77048 & 20.839 & 0.049 & 2453045.83694 & 20.521 & 0.092 \\
cep058 & 2453055.62418 & 20.608 & 0.059 & 2453055.70117 & 20.522 & 0.068 \\
cep059 & 2453045.77048 & 20.901 & 0.043 & 2453045.83694 & 20.475 & 0.083 \\
cep059 & 2453055.62418 & 20.644 & 0.063 & 2453055.70117 & 20.210 & 0.039 \\
cep060 &       --      &   --   &  --   & 2453036.80998 & 20.225 & 0.048 \\
cep060 & 2453047.68199 & 20.927 & 0.059 & 2453047.76282 & 20.494 & 0.074 \\
cep061 &       --      &   --   &  --   & 2453036.80998 & 19.392 & 0.073 \\
cep061 & 2453047.68199 & 19.683 & 0.041 & 2453047.76282 & 19.241 & 0.047 \\
cep063 & 2453045.77048 & 20.395 & 0.049 & 2453045.83694 & 20.252 & 0.133 \\
cep064 & 2453037.69282 & 20.832 & 0.039 & 2453037.76573 & 20.255 & 0.057 \\
cep064 & 2453049.67836 & 20.594 & 0.075 &       --      &   --   &  --   \\
cep066 & 2453037.69282 & 20.800 & 0.046 & 2453037.76573 & 20.152 & 0.070 \\
cep066 & 2453045.77048 & 20.513 & 0.062 & 2453045.83694 & 20.028 & 0.072 \\
cep066 & 2453049.67836 & 20.576 & 0.136 & 2453049.74253 & 20.658 & 0.044 \\
cep066 & 2453055.62418 & 20.603 & 0.073 & 2453055.70117 & 20.376 & 0.055 \\
cep067 & 2453037.69282 & 21.140 & 0.043 & 2453037.76573 & 20.601 & 0.065 \\
cep067 & 2453049.67836 & 20.973 & 0.160 & 2453049.74253 & 20.604 & 0.048 \\
cep067 & 2453055.62418 & 21.029 & 0.083 & 2453055.70117 & 20.616 & 0.077 \\
cep068 & 2453037.69282 & 20.498 & 0.034 & 2453037.76573 & 20.120 & 0.053 \\
cep068 & 2453049.67836 & 20.665 & 0.105 & 2453049.74253 & 20.538 & 0.042 \\
cep069 & 2453045.77048 & 19.927 & 0.029 & 2453045.83694 & 19.158 & 0.033 \\
cep069 & 2453055.62418 & 20.084 & 0.038 & 2453055.70117 & 19.222 & 0.036 \\
cep072 & 2453045.77048 & 20.475 & 0.039 & 2453045.83694 & 20.140 & 0.068 \\
cep072 & 2453055.62418 & 20.980 & 0.065 & 2453055.70117 & 20.381 & 0.063 \\
cep073 & 2453045.77048 & 20.647 & 0.045 & 2453045.83694 & 20.206 & 0.063 \\
cep073 & 2453055.62418 & 20.891 & 0.078 & 2453055.70117 & 20.554 & 0.049 \\
cep074 & 2453037.69282 & 20.781 & 0.053 & 2453037.76573 & 20.499 & 0.083 \\
cep074 & 2453049.67836 & 20.818 & 0.096 & 2453049.74253 & 20.421 & 0.062 \\
cep075 & 2453037.69282 & 21.113 & 0.052 & 2453037.76573 & 20.732 & 0.082 \\
cep075 & 2453049.67836 & 21.136 & 0.114 & 2453049.74253 & 20.663 & 0.057 \\
cep076 &       --      &   --   &  --   & 2453036.80998 & 20.381 & 0.062 \\
cep076 & 2453047.68199 & 20.820 & 0.061 & 2453047.76282 & 20.581 & 0.081 \\
cep077 & 2453037.69282 & 21.029 & 0.050 & 2453037.76573 & 20.431 & 0.076 \\
cep077 & 2453049.67836 & 21.142 & 0.118 & 2453049.74253 & 20.484 & 0.080 \\
cep078 & 2453045.77048 & 21.156 & 0.092 & 2453045.83694 & 20.947 & 0.164 \\
cep078 & 2453055.62418 & 21.153 & 0.084 & 2453055.70117 & 20.692 & 0.066 \\
cep079 & 2453037.69282 & 21.279 & 0.069 & 2453037.76573 & 20.423 & 0.080 \\
cep079 & 2453049.67836 & 20.727 & 0.162 & 2453049.74253 & 20.685 & 0.047 \\
cep080 & 2453045.77048 & 20.688 & 0.054 & 2453045.83694 & 20.466 & 0.082 \\
cep080 & 2453055.62418 & 20.945 & 0.096 & 2453055.70117 & 20.592 & 0.074 \\
cep084 & 2453047.68199 & 20.877 & 0.085 & 2453047.76282 & 20.569 & 0.118 \\
cep088 & 2453037.69282 & 19.652 & 0.027 & 2453037.76573 & 18.678 & 0.024 \\
cep088 & 2453049.67836 & 19.633 & 0.030 & 2453049.74253 & 18.606 & 0.023 \\
cep090 & 2453037.69282 & 20.940 & 0.051 & 2453037.76573 & 20.441 & 0.073 \\
cep090 & 2453049.67836 & 21.047 & 0.094 & 2453049.74253 & 20.390 & 0.058 \\
cep092 & 2453045.77048 & 21.017 & 0.043 & 2453045.83694 & 20.572 & 0.087 \\
cep092 & 2453055.62418 & 21.314 & 0.066 & 2453055.70117 & 20.473 & 0.072 \\
cep094 & 2453037.69282 & 21.130 & 0.041 & 2453037.76573 & 20.576 & 0.076 \\
cep094 & 2453049.67836 & 21.284 & 0.138 & 2453049.74253 & 20.895 & 0.060 \\
cep095 &       --      &   --   &  --   & 2453036.80998 & 19.765 & 0.051 \\
cep095 & 2453047.68199 & 20.420 & 0.050 & 2453047.76282 & 19.702 & 0.061 \\
cep096 &       --      &   --   &  --   & 2453036.80998 & 20.633 & 0.077 \\
cep096 & 2453047.68199 & 20.476 & 0.055 &       --      &   --   &  --   \\
cep097 & 2453045.77048 & 21.048 & 0.052 & 2453045.83694 & 20.725 & 0.098 \\
cep097 & 2453055.62418 & 20.892 & 0.066 & 2453055.70117 & 20.413 & 0.054 \\
cep098 &       --      &   --   &  --   & 2453036.80998 & 19.728 & 0.045 \\
cep098 & 2453047.68199 & 20.296 & 0.038 & 2453047.76282 & 19.874 & 0.052 \\
cep099 & 2453045.77048 & 21.000 & 0.061 & 2453045.83694 & 20.377 & 0.080 \\
cep099 & 2453055.62418 & 20.915 & 0.094 & 2453055.70117 & 20.785 & 0.058 \\
cep100 & 2453045.77048 & 21.580 & 0.084 & 2453045.83694 & 20.691 & 0.105 \\
cep100 & 2453055.62418 & 21.695 & 0.092 & 2453055.70117 & 20.969 & 0.098 \\
cep101 & 2453037.69282 & 21.250 & 0.072 & 2453037.76573 & 20.610 & 0.094 \\
cep103 & 2453045.77048 & 21.235 & 0.066 & 2453045.83694 & 20.801 & 0.108 \\
cep103 & 2453055.62418 & 21.015 & 0.104 & 2453055.70117 & 21.059 & 0.064 \\
cep104 &       --      &   --   &  --   & 2453045.83694 & 19.336 & 0.039 \\
cep104 & 2453055.62418 & 20.661 & 0.042 & 2453055.70117 & 19.187 & 0.265 \\
cep105 &       --      &   --   &  --   & 2453036.80998 & 20.525 & 0.089 \\
cep105 & 2453047.68199 & 20.682 & 0.067 &       --      &   --   &  --   \\
cep108 & 2453045.77048 & 21.395 & 0.063 &       --      &   --   &  --   \\
cep108 & 2453055.62418 & 21.586 & 0.122 & 2453055.70117 & 21.143 & 0.082 \\
cep110 &       --      &   --   &  --   & 2453036.80998 & 20.590 & 0.085 \\
cep110 & 2453047.68199 & 20.874 & 0.063 & 2453047.76282 & 20.334 & 0.084 \\
\enddata
\end{deluxetable}

\clearpage
\begin{deluxetable}{lcccccc}
\tablecolumns{7}
\tablewidth{0pc}
\tablecaption{Intensity mean $J$ and $K$ magnitudes for 77 Cepheids in NGC~3109}
\tablehead{
\multicolumn{1}{c}{ID} & $P$ & $\langle{J}\rangle$ & $\sigma_{\langle{J}\rangle}$ & $\langle{K}\rangle$ & $\sigma_{\langle{K}\rangle}$ & Remarks}
\startdata
cep001 & 31.4793 & 18.864 & 0.033 & 18.413 & 0.028 & \\
cep002 & 31.270  & 18.825 & 0.039 & 18.242 & 0.030 & \\
cep003 & 29.110  & 19.077 & 0.038 & 18.544 & 0.034 & \\
cep004 & 27.389  & 18.950 & 0.033 & 18.502 & 0.030 & \\
cep005 & 26.8274 & 19.068 & 0.037 & 18.551 & 0.040 & \\
cep007 & 20.388  & 19.795 & 0.035 & 19.252 & 0.039 & \\
cep009 & 19.5759 & 19.328 & 0.035 & 18.861 & 0.031 & \\
cep011 & 17.2293 & 19.767 & 0.036 & 19.261 & 0.039 & \\
cep012 & 14.750  & 19.812 & 0.045 & 19.265 & 0.049 & \\
cep014 & 14.062  & 19.650 & 0.048 & 19.282 & 0.039 & \\
cep015 & 14.030  & 19.949 & 0.040 & 19.375 & 0.040 & \\
cep016 & 13.9047 & 19.648 & 0.039 & 19.229 & 0.038 & \\
cep017 & 13.6250 & 19.890 & 0.039 & 19.418 & 0.040 & \\
cep018 & 13.364  & 20.189 & 0.063 & 19.675 & 0.051 & \\
cep020 & 13.011  & 20.230 & 0.039 & 19.757 & 0.046 & \\
cep022 & 11.864  & 19.836 & 0.053 & 19.520 & 0.050 & \\
cep023 & 11.707  & 19.660 & 0.040 & 19.166 & 0.034 & \\
cep025 & 11.596  & 18.600 & 0.031 & 17.769 & 0.028 & blend? \\
cep026 & 11.534  & 20.423 & 0.054 & 19.949 & 0.061 & \\
cep027 & 11.238  & 20.108 & 0.047 & 19.641 & 0.052 & \\
cep028 & 11.0663 & 20.295 & 0.048 & 19.715 & 0.049 & \\
cep029 & 10.903  & 19.798 & 0.044 & 19.306 & 0.040 & \\
cep030 & 10.887  & 20.198 & 0.070 & 19.799 & 0.046 & \\
cep031 & 10.851  & 20.025 & 0.038 & 19.492 & 0.043 & \\
cep032 & 10.826  & 20.316 & 0.054 & 19.520 & 0.064 & \\
cep035 & 9.7501  & 20.071 & 0.058 & 19.773 & 0.066 & \\
cep036 & 9.4084  & 20.270 & 0.069 & 19.957 & 0.051 & \\
cep038 & 9.3242  & 20.408 & 0.053 & 20.011 & 0.053 & \\
cep039 & 9.11136 & 20.608 & 0.062 & 20.291 & 0.060 & \\
cep043 & 8.7970  & 20.306 & 0.052 & 20.215 & 0.075 & \\
cep047 & 8.55708 & 20.262 & 0.057 & 19.833 & 0.060 & \\
cep048 & 8.4791  & 20.530 & 0.062 & 19.980 & 0.052 & \\
cep050 & 8.24330 & 20.592 & 0.071 & 20.149 & 0.060 & \\
cep051 & 8.19173 & 20.597 & 0.052 & 20.352 & 0.065 & \\
cep052 & 8.10451 & 20.541 & 0.057 & 20.251 & 0.060 & \\
cep053 & 7.9351  & 20.651 & 0.062 & 20.301 & 0.073 & \\
cep054 & 7.9038  & 20.456 & 0.053 & 20.131 & 0.064 & \\
cep055 & 7.7960  & 20.706 & 0.066 & 20.105 & 0.066 & \\
cep056 & 7.77376 & 19.853 & 0.044 & 18.717 & 0.044 & blend? \\
cep057 & 7.5389  & 19.834 & 0.050 & 18.952 & 0.071 & blend? \\
cep058 & 7.4089  & 20.721 & 0.058 & 20.507 & 0.084 & \\
cep059 & 7.3858  & 20.736 & 0.058 & 20.296 & 0.068 & \\
cep060 & 7.2814  & 20.722 & 0.066 & 20.339 & 0.066 & \\
cep061 & 7.25995 & 19.699 & 0.051 & 19.318 & 0.065 & blend? \\
cep063 & 7.15940 & 20.490 & 0.057 & 20.304 & 0.136 & \\
cep064 & 7.13061 & 20.767 & 0.063 & 20.297 & 0.064 & \\
cep066 & 6.8511  & 20.586 & 0.088 & 20.278 & 0.063 & \\
cep067 & 6.8294  & 21.114 & 0.108 & 20.688 & 0.067 & \\
cep068 & 6.6932  & 20.647 & 0.081 & 20.343 & 0.052 & \\
cep069 & 6.6735  & 19.989 & 0.040 & 19.183 & 0.041 & blend? \\
cep072 & 6.4250  & 20.742 & 0.058 & 20.263 & 0.069 & \\
cep073 & 6.3065  & 20.809 & 0.067 & 20.395 & 0.060 & \\
cep074 & 6.2719  & 20.915 & 0.080 & 20.517 & 0.076 & \\
cep075 & 6.2713  & 20.806 & 0.091 & 20.425 & 0.074 & \\
cep076 & 6.16368 & 20.896 & 0.068 & 20.517 & 0.075 & \\
cep077 & 6.1329  & 21.025 & 0.093 & 20.438 & 0.081 & \\
cep078 & 6.1098  & 21.151 & 0.091 & 20.858 & 0.127 & \\
cep079 & 6.0795  & 20.861 & 0.126 & 20.471 & 0.069 & \\
cep080 & 6.0749  & 20.919 & 0.081 & 20.582 & 0.081 & \\
cep084 & 5.9343  & 21.042 & 0.090 & 20.730 & 0.122 & \\
cep088 & 5.5369  & 19.677 & 0.036 & 18.680 & 0.032 & \\
cep090 & 5.4844  & 21.052 & 0.079 & 20.450 & 0.069 & \\
cep092 & 5.39909 & 21.227 & 0.060 & 20.595 & 0.083 & \\
cep094 & 5.2442  & 21.317 & 0.104 & 20.780 & 0.072 & \\
cep095 & 5.2420  & 20.262 & 0.058 & 19.639 & 0.060 & \\
cep096 & 5.2163  & 20.620 & 0.063 & 20.670 & 0.083 & \\
cep097 & 5.17264 & 20.922 & 0.063 & 20.584 & 0.082 & \\
cep098 & 5.0493  & 20.405 & 0.048 & 19.831 & 0.053 & \\
cep099 & 5.03650 & 20.800 & 0.082 & 20.452 & 0.073 & \\
cep100 & 4.9883  & 21.523 & 0.091 & 20.776 & 0.104 & \\
cep101 & 4.9221  & 21.259 & 0.078 & 20.557 & 0.099 & \\
cep102 & 4.7765  &   --   &   --  & 20.980 & 0.125 & \\
cep103 & 4.5808  & 21.053 & 0.090 & 20.852 & 0.091 & \\
cep104 & 4.5658  & 20.758 & 0.052 & 19.281 & 0.191 & \\
cep105 & 4.5120  & 20.787 & 0.073 & 20.466 & 0.094 & \\
cep108 & 4.0877  & 21.589 & 0.099 & 21.281 & 0.087 & \\
cep110 & 3.9641  & 20.776 & 0.070 & 20.399 & 0.087 & 
\enddata
\end{deluxetable}

\begin{deluxetable}{lclll}
\tablecolumns{5}
\tablewidth{0pc}
\tablecaption{Previous and present distance determinations to NGC 3109}
\tablehead{
\multicolumn{1}{c}{Reference} & $(m-M)_0$ & Errors & Method & Wavebands }
\startdata
\citet{dik85}  &  25.98  & $\pm0.15$ &  Cepheids      & {\it VR} (photographic) \\
\citet{ef85}   &  25.34  & $\pm0.1$ (random) &  M supergiants & {\it JHK} \\
\citet{sc88}   &  26.0~~ &           &  Cepheids      & {\it B} (photographic) \\
\citet{cpb92}  &  25.5~~ & $\pm0.2$  &  Cepheids      & {\it BV} \\
\citet{rm92}   &  25.96  & $+0.1\atop-0.4$  &  planetary nebulae   & narrow bandwidth \\
\citet{lee93}  &  25.45  & $\pm0.15$ &  TRGB          & {\it I} \\
\citet{mpc97}  &  25.67  & $\pm0.16$ &  Cepheids      & {\it BVRI} \\
\citet{mza99}  &  25.62  & $\pm0.1$  &  TRGB          & {\it I} \\
\citet{kar02}  &  25.62  & $\pm0.13$ & TRGB & {\it I} \\
\citet{men02}  &  25.52  & $\pm0.06$ (internal) & TRGB & {\it I} \\
               &         & $\pm0.18$ (systematic) & \\
this paper     &  25.57  & $\pm0.02$ (random) & Cepheids & {\it VIJK} \\
               &         & $\pm0.06$ (systematic) & \\
\enddata
\end{deluxetable}

\clearpage
\begin{figure}
\includegraphics[width=16cm]{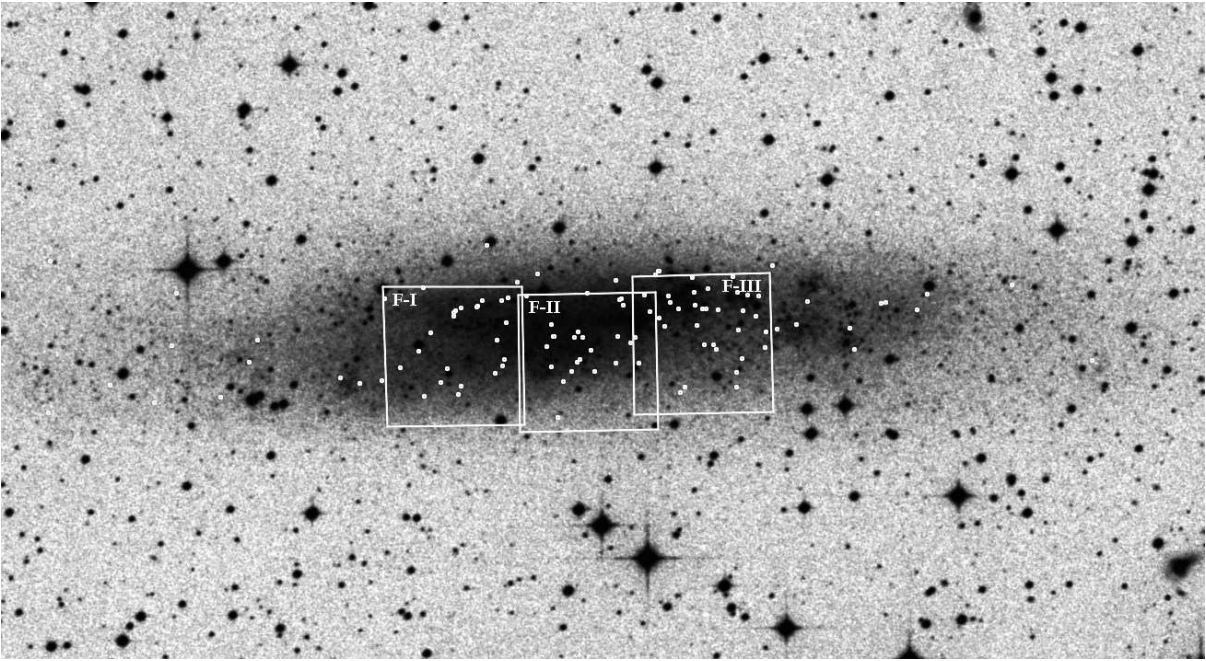}
\caption{Contours of the observed fields in NGC~3109 on the DSS-2 red
  plate. White points mark the Cepheids reported in Paper~I.}
\end{figure}

\begin{figure}
\vspace{-1cm}
\includegraphics[width=16cm]{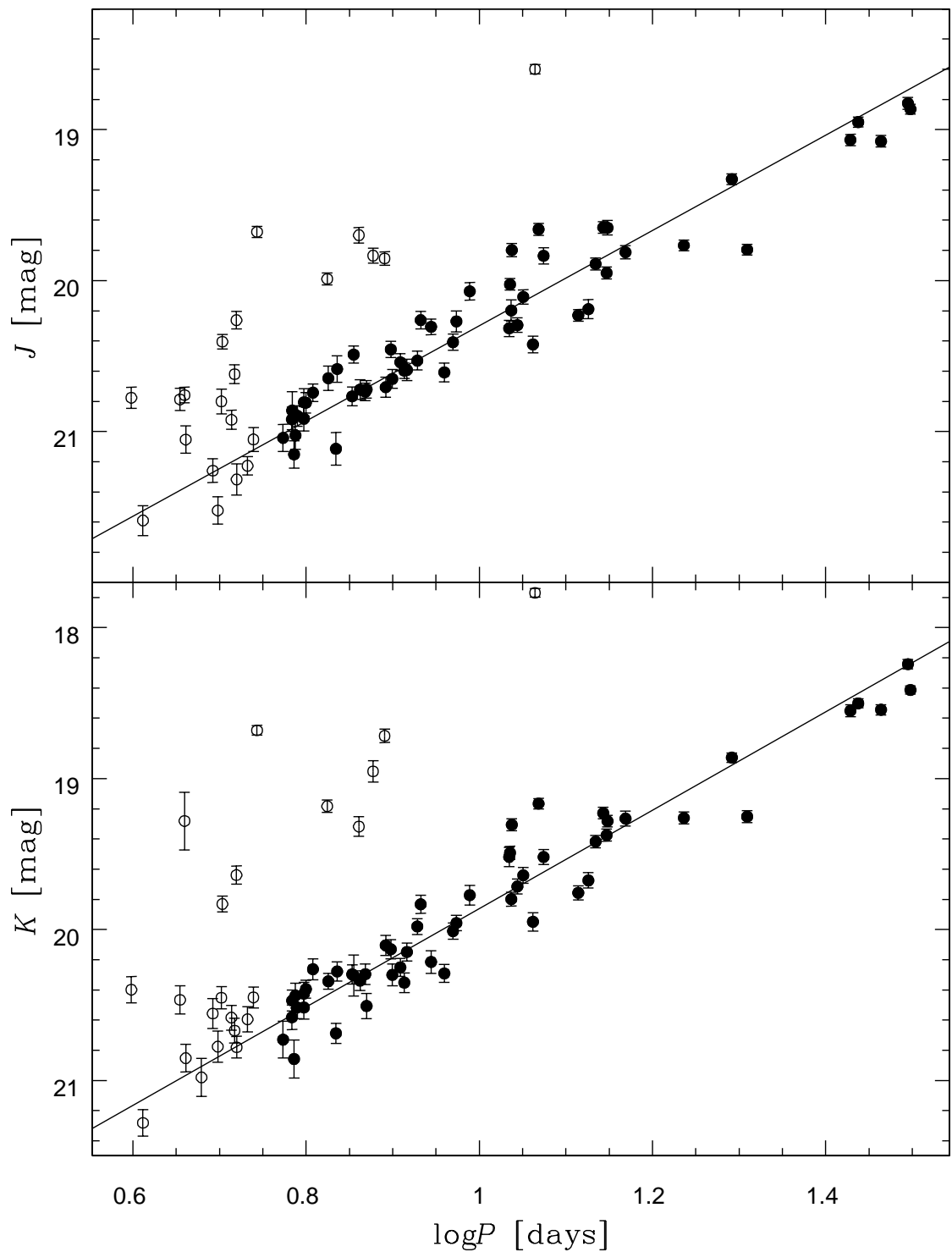}
\vspace{-0.5cm}
\caption{$J$- and $K$-band {\it PL} relations for Cepheids in NGC~3109. The
  straight lines are the linear least-squares fits to the points marked with
  filled circles. In both diagrams, we adopted the slopes derived by Persson
  et al. for the LMC Cepheids. The empty circles are short-period and
  outlying Cepheids omitted from the fit (see text).}
\end{figure}

\begin{figure}
\includegraphics[width=16cm]{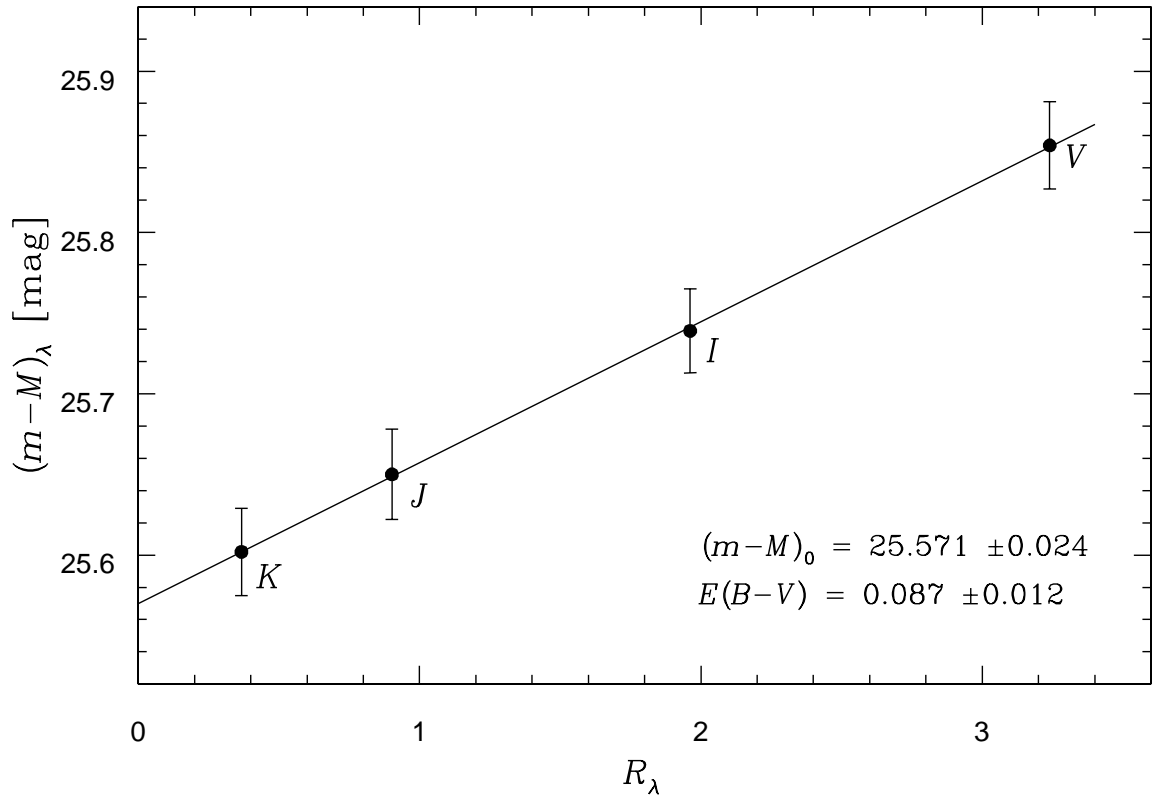}
\vspace{-9.5cm}
\caption{Apparent distance moduli to NGC~3109 determined in different
  photometric bands, versus the ratio of total to selective absorption for
  these bands. The intersection and slope of the best-fitting line gives the
  true distance modulus and the total color excess, respectively.}
\end{figure}

\begin{figure}
\includegraphics[width=16cm]{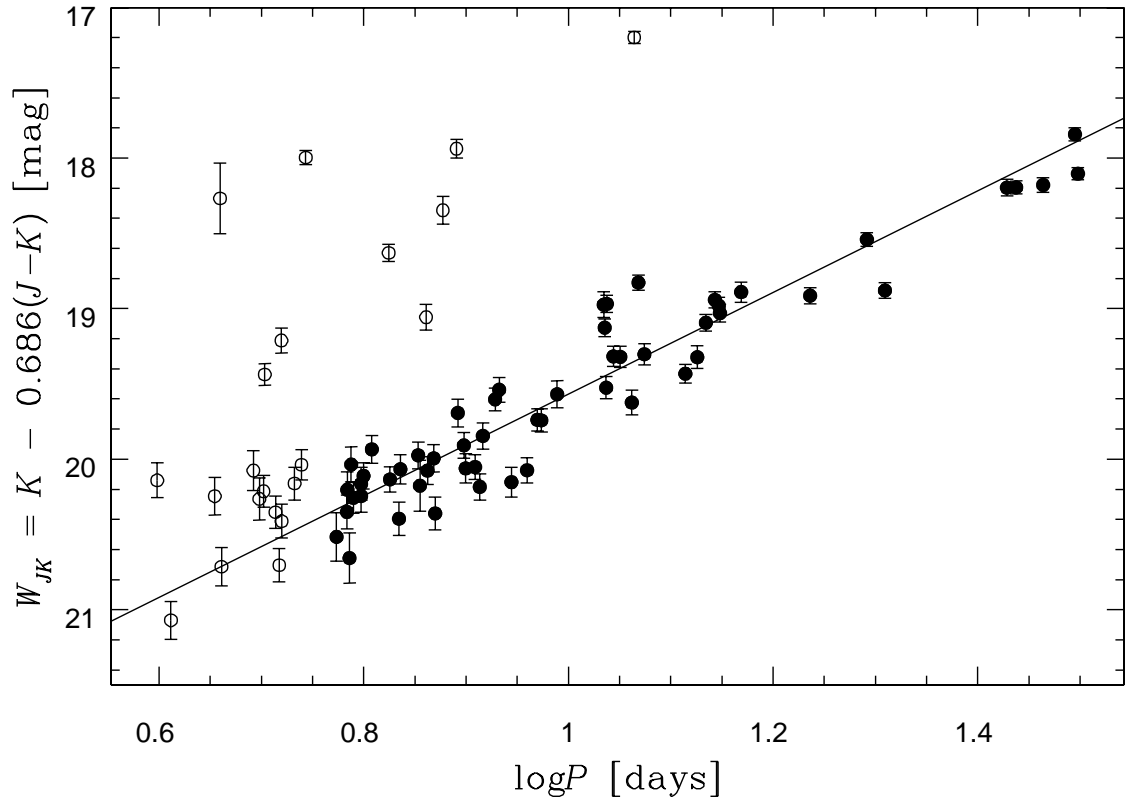}
\vspace{-9.5cm}
\caption{The period -- NIR Wesenheit index diagram for Cepheids in NGC~3109.
  The line shows the linear least-squares fit to the points indicated by
  filled circles, adopting the slope derived from the LMC Cepheids. The empty
  circles are short-period and outlying Cepheids omitted from the fit.}
\end{figure}

\begin{figure}
\includegraphics[width=16cm]{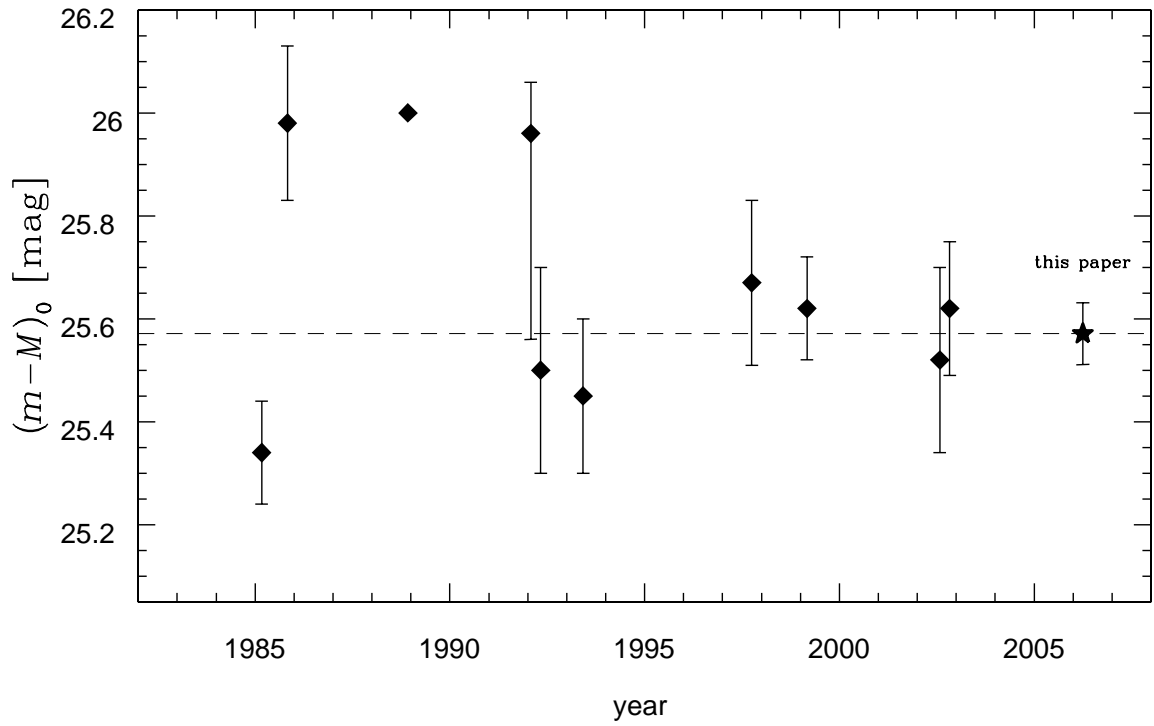}
\vspace{-11cm}
\caption{Previous distance determinations to NGC 3109 (diamonds) in comparison
  to our present determination (star). Numerical values and techniques used
  for the individual determinations are given in Table 3.}
\end{figure}

\end{document}